\documentclass[aps,prb,twocolumn,amsmath,superscriptaddress]{revtex4-2}
\usepackage{graphicx}
\usepackage{hyperref}
\usepackage{mathrsfs}
\hypersetup{hypertex=true,
	colorlinks=true,
	anchorcolor=blue,
	linkcolor=blue,
    citecolor=blue}

\begin{document}

\title{Charge and spin transport through normal lead coupled to s-wave superconductor and a Majorana zero mode}

\author{Yue Mao}
\affiliation{International Center for Quantum Materials, School of Physics, Peking University, Beijing 100871, China}
\author{Qing-Feng Sun}
\email[]{sunqf@pku.edu.cn}
\affiliation{International Center for Quantum Materials, School of Physics, Peking University, Beijing 100871, China}
\affiliation{Collaborative Innovation Center of Quantum Matter, Beijing 100871, China}
\affiliation{CAS Center for Excellence in Topological Quantum Computation, University of Chinese Academy of Sciences, Beijing 100190, China
}

\date{\today}

\begin{abstract}
	
Zero-bias charge conductance peak (ZBCCP) is a significant symbol of Majorana zero modes (MZMs).
The proximity effect of s-wave superconductor is usually demanded
in the fabrication of MZMs. So in transport experiments,
the system is inevitably coupled to the s-wave superconductor.
Here we study how the ZBCCP is affected by coupling of the s-wave superconductor.
The results show that the ZBCCP could be changed into
a zero-bias valley due to the coupling of s-wave superconductor,
although the conductance at the zero bias still keeps a quantized value $2e^2/h$.
So it does not mean no MZM exists when no ZBCCP is experimentally observed.
In addition, the spin transport is investigated. Four reflection processes
(the normal reflection, spin-flip reflection, normal Andreev reflection,
and equal-spin Andreev reflection) usually occur. The reflection coefficients
are strongly dependent on the spin direction of the incident electron,
and they may be symmetrical or Fano resonance shapes.
But the spin conductance always shows a zero-bias peak with the height $e/2\pi$
regardless of the direction of the spin bias
and the coupling strength of the s-wave superconductor.
So measuring spin transport properties could be a more reliable method
to judge the existence of MZMs.

\end{abstract}

\maketitle

\section{\label{SEC1} Introduction}
Majorana fermion is a fundamental particle which is its own antiparticle.
It obeys non-Abelian statistics and has attracted much attention
as potential applications in fault tolerant topological quantum
computation.\cite{Kitaev_TQC, Nayak_TQC}
In some condensed systems the excitation of quasiparticles has been discovered to have
the characteristics of Majorana fermions and is called Majorana zero mode
(MZM).\cite{Elliott_MZM_RMP, Lutchyn_MZMT, Oreg_MZMT, Mourik_MZME, Das_MZME, Chen_MZME, Nichele_MZME, Zhang_MZME, Lutchyn_MZME, Vaitiekenas_MZME, Nadj-Perge_Fe_Chain, Fu_Proximity, Wang_MZM_TS, Xu_MZM_TS1, Xu_MZM_TS2, Sun_MZM_TS3, Zhang_iron-base1, Wang_iron-base2, Zhu_iron-base3}
So far, MZMs are predicted and/or realized in several setups,
including semiconductor nanowire under a high magnetic field and
in proximity to a superconductor,\cite{Lutchyn_MZMT, Oreg_MZMT, Mourik_MZME, Das_MZME, Chen_MZME, Nichele_MZME, Zhang_MZME, Lutchyn_MZME, Vaitiekenas_MZME} ferromagnetic atomic chains on a superconductor,\cite{Nadj-Perge_Fe_Chain}
topological insulator-superconductor heterostructure,\cite{Fu_Proximity, Wang_MZM_TS, Xu_MZM_TS1, Xu_MZM_TS2, Sun_MZM_TS3}
iron-based superconductor,\cite{Zhang_iron-base1, Wang_iron-base2, Zhu_iron-base3}
organic material,\cite{Tang_DNA} and so on.

Zero-bias charge conductance peak (ZBCCP) is a significant symbol of MZMs.
By solving a model of normal lead coupled to a MZM,
it is theoretically predicted that a resonant Andreev reflection happens
due to the coupling strength between the electron and MZM being exactly equal
to that between the hole and MZM, and the ZBCCP's height is quantized at $2e^2/h$.\cite{Law_ZBP}
It provides a simple but effective method for detections of MZMs,
which is widely used in many experiments.\cite{Mourik_MZME, Das_MZME, Chen_MZME, Nichele_MZME, Zhang_MZME,Lutchyn_MZME, Nadj-Perge_Fe_Chain, Wang_MZM_TS, Xu_MZM_TS1, Xu_MZM_TS2, Sun_MZM_TS3, Wang_iron-base2, Zhu_iron-base3, Vaitiekenas_MZME}
Nevertheless, this ZBCCP can also be caused by Kondo effect,
disorder, weak antilocalization, Andreev bound states
and so on.\cite{Lee_QITA, Liu_QITA, Rainis_QITA, Lee_QITA2, Pikulin_QITA}
As a result, this feature does not completely determine the existence of MZMs.
To distinguish these factors and get more unambiguous evidences
for MZMs still remains challenging.\cite{Vuik_ZBV}

Spintronics is an energetic subject developed in last two
decades.\cite{Prinz_Spin1, Wolf_Spin2, Fert_Spin3}
It has extensive applying future because of its
superior storage performance and low-power consumption.
Spin transport is the counterpart to charge transport.
Charge bias gives rise to the difference of the chemical potentials of
the left and right leads, while spin bias splits
the chemical potentials of electrons with opposite directions of spin.\cite{Wang_Spin_bias}
Under a spin bias, spin-up and spin-down electrons move oppositely
and a spin current is induced.
Spin is a vector, so spin bias is vector and spin current is tensor,
which are different from charge bias and charge current.\cite{Shi_Spin_current, Sun_Spin_current}
Experimentally spin current has been observed widely and
is of significance to characterization of quantum
materials.\cite{Stevens_Spin_current, Han_Spin_current}
A MZM leads to the selective equal-spin Andreev reflection
by reflecting an electron into an equal-spin hole,
transmitting a spin polarized current.\cite{He_AR}
Therefore, the Majorana junction could response to a spin bias and produce a spin current.

In the transport experimental device, MZM is bedded on p-wave component of superconductors which consists in
a material with strong spin-orbit coupling and
ferromagnetic exchange field (or magnetic field)
in proximity to an s-wave superconductor. This s-wave superconductor could play an important role.\cite{Rainis_QITA}
When a normal lead is coupled to MZMs,
it is usually coupled to the s-wave superconductor too.
Furthermore, because of the spin-orbit coupling and magnetic field
in the Majorana system, spin is not a good quantum number,
and the Cooper pairs have both spin-singlet and spin-triplet components.
So this system has both s-wave and p-wave components of superconductors,
implying that the normal lead is unavoidably coupled to the s-wave superconductor,
while coupling to the MZM.
On this condition, an incident electron from the normal lead is
reflected by both MZM and s-wave superconductor.
What happens to the transport phenomenon?
Can ZBCCP still exist with the coupling of the s-wave superconductor?

\begin{figure}
	\includegraphics[width=\columnwidth]{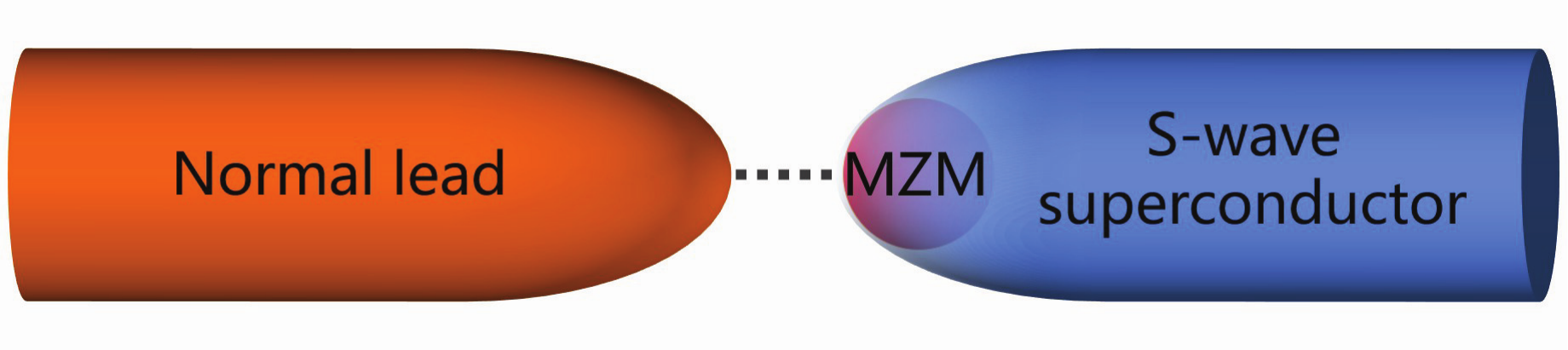}%
	\caption{\label{FIG1} The schematic diagram of the considered device which
consists of a normal lead coupling to a MZM and an s-wave superconductor.}
\end{figure}

In this paper, we study the transport properties of
a normal lead coupled to a MZM and an s-wave superconductor,
as shown in Fig.~\ref{FIG1}.
We find that, the coupling to s-wave superconductor could change the ZBCCP
to a zero-bias charge conductance valley.
So a MZM does not necessarily induce a ZBCCP.
In addition, owing to the nontopological explanations of ZBCCP,
this peak-shape conductance curve is also insufficient
to define the MZM.\cite{Lee_QITA, Liu_QITA, Rainis_QITA, Lee_QITA2, Pikulin_QITA}
Therefore, there is no direct connection between MZM and ZBCCP.
On the contrary, the spin conductance curve is always a zero-bias peak
with a zero-bias quantized value $ e/2\pi$.
Its peak shape is robust against the coupling strength to s-wave superconductor and
the direction of the spin bias.

The rest of this paper is as follows:
In Sec.~\ref{SEC2}, we present the Hamiltonian of
the system consisting of normal lead coupled
to MZM and superconductor and derive the expressions of reflection coefficients
and charge (spin) conductance. In Sec.~\ref{SEC3} and Sec.~\ref{SEC4}, we show the results of charge and spin transport respectively and
give analysis and discussion. At last, a conclusion is given in Sec.~\ref{SEC5}.

\section{\label{SEC2} Model and formula}
We consider a system consisting of a normal lead coupled to a MZM and an s-wave superconductor.
The general Hamiltonian is
\begin{equation}
	H=H_L+H_S+H_{LS}+H_{L\gamma},
\end{equation}
where $ H_L $, $ H_S $, $ H_{LS} $, and $ H_{L\gamma}$
are the Hamiltonians of normal lead, s-wave superconductor,
coupling between normal lead and superconductor,
and coupling between normal lead and MZM, respectively:\cite{Liu_Hamiltonian1,Ricco_Hamiltonian2,Gong_Hamiltonian3}
\begin{eqnarray}
	H_L&=&\sum_{k\sigma} \epsilon_k^L d_{k\sigma}^\dagger d_{k\sigma}, \label{HLL}\\
	H_S&=&\sum_{k\sigma} \epsilon_k^S c_{k\sigma}^\dagger c_{k\sigma} + \sum_{k} (\Delta c_{k\uparrow} c_{-k\downarrow} + \Delta c_{-k\downarrow}^\dagger c_{k\uparrow}^\dagger),
     \label{HSS} \\
	H_{LS}&=&\sum_{k k^{'} \sigma} t_S d_{k\sigma}^{ \dagger} c_{k^{'}\sigma} +h.c. ,\\
	H_{L\gamma}&=&\sum_{k} it(d_{kz} + d_{kz}^\dagger) \gamma . \label{HLg}
\end{eqnarray}
Here $d_{k\sigma}^\dagger$ ($d_{k\sigma}$) and $c_{k\sigma}^\dagger$ ($c_{k\sigma}$)
with $\sigma=\uparrow,\downarrow$ are the creation (annihilation)
operators of electrons in the normal lead and s-wave superconductor.
$\gamma$ describes the operator of the MZM.
$\Delta$ is the superconducting gap,
and $t_S$ ($t$) is the coupling strength between the normal lead and s-wave superconductor
(MZM).
In Eqs.(\ref{HLL} and \ref{HSS}),
the normal lead and s-wave superconductor are considered to
have only a single channel, and the multi-channel effects
are ignored.\cite{Beenakker_Multi-channel}
Because of the self-Hermitian property of MZM $\gamma^\dagger=\gamma $,
the MZM is actually just coupled to electrons with spin pointing to a certain direction.\cite{He_AR}
In Eq.(\ref{HLg}), we assume that the MZM only couples to electrons with $ +z $-direction spin
in the normal lead.

In order to solve the reflection coefficients for the incident electrons with spin of any direction,
we rotate the spinor by a unitary transformation without loss of generality
\begin{equation}
	\begin{pmatrix}
		d_{kz} \\
		d_{k \bar{z}} \\
	\end{pmatrix}
	=\begin{pmatrix}
		\cos \frac{\theta}{2} & \sin \frac{\theta}{2} e^{i \phi} \\
		-\sin \frac{\theta}{2} e^{-i \phi} & \cos \frac{\theta}{2} \\
	\end{pmatrix}
	\begin{pmatrix}
		d_{k\hat{n}} \\
		d_{k\hat{\bar{n}}} \\
	\end{pmatrix}.
\end{equation}
Here the subscripts $z$, $\bar{z}$, $\hat{n}$, and $\hat{\bar{n}}$ respectively represent
the spin with $+z$, $-z$, $\hat{n}$, and $-\hat{n}$ directions,
with $ \hat{n} = (\sin \theta \cos \phi, \sin \theta \sin \phi, \cos \theta) $
and $\theta $ and $ \phi $ being Euler angles of spherical coordinates.
Note that Hamiltonians $ H_L $, $ H_S $, and $ H_{LS} $ have isotropy of spin, so the subscript
$\sigma=\uparrow, \downarrow $ could indicate any spin direction.
Setting $ \uparrow=\hat{n} $ is helpful to simplify the problem.

Let us consider a small charge or spin bias applied to the system,
with the bias smaller than the superconducting gap.
In this case the transmitting process from the normal lead
to the superconductor can not happen.
Using the Landauer-B$\ddot{u}$ttiker formula,
the particle currents with spin $ \hat{n} $ is \cite{Meir_GS, Sun_GS, Cheng_GS}
\begin{align}
I_{\hat{n}} & = \frac{1}{h} \int_{-\infty}^{+\infty} dE [
S_{e \hat{\bar{n}} e \hat{n}} ( f_{Le \hat{n}} - f_{Le \hat{\bar{n}}} )  \notag\\
 &+ S_{h \hat{n} e \hat{n}} ( f_{Le \hat{n}} - f_{Lh \hat{n}} )
+ S_{h \hat{\bar{n}} e \hat{n}} ( f_{Le \hat{n}} - f_{Lh \hat{\bar{n}}} ) ],\label{In}
\end{align}
where $ f_{L e/h, \hat{n}/\hat{\bar{n}}} =1/\{exp[(E \mp \mu_{\hat{n}/\hat{\bar{n}}})/{k_B T}]+1\} $
are the Fermi distribution functions with the temperature $T$
and the chemical potentials $\mu_{\hat{n}}$ and $\mu_{\hat{\bar{n}}}$.
These $S$'s in Eq.(\ref{In}) are the reflection coefficients.
There are totally 4 reflection coefficients for the incident electron with spin $\hat{n}$:
$S_{e \hat{n} e \hat{n}} $ and $ S_{e \hat{\bar{n}} e \hat{n}} $
are the normal reflection and spin-flip reflection,
and $S_{h \hat{\bar{n}} e \hat{n}} $ and $ S_{h \hat{n} e \hat{n}} $
are the normal Andreev reflection and equal-spin Andreev reflection.

If a charge bias $ V_c $ is applied, $ \mu_{\hat{n}}=\mu_{\hat{\bar{n}}}=eV_c $
and charge current is $ I_c=e(I_{\hat{n}}+I_{\hat{\bar{n}}}) $.
As for a $ \hat{n} $-direction spin bias $ V_s $,
$ \mu_{\hat{n}}=  -\mu_{\hat{\bar{n}}} = eV_s $
and spin current is $ I_{s\hat{n}}=\frac{\hbar}{2}(I_{\hat{n}} - I_{\hat{\bar{n}}})
$.\cite{Wang_Spin_bias,Sun_Spin_bias}
The charge (spin) conductance under a charge (spin) bias is $ G_c=dI_c/dV_c $
($ G_{s\hat{n}}=dI_{s\hat{n}}/dV_s $)
and is simplified into at zero temperature
\begin{widetext}
\begin{eqnarray}
	G_c&=&\frac{e^2}{h}
 \sum_{\epsilon \in \{eV_c, -eV_c\}}
 [S_{h\hat{n} e\hat{n}}(\epsilon)+S_{h\hat{\bar{n}} e\hat{n}}(\epsilon)
  +S_{h\hat{n} e\hat{\bar{n}}}(\epsilon)+S_{h\hat{\bar{n}} e\hat{\bar{n}}}(\epsilon)
  ],\label{GC1} \\
  	G_{s\hat{n}}&=& \frac{e}{4\pi}
    \sum_{\epsilon \in \{eV_s, -eV_s\}}
  [S_{h\hat{n} e\hat{n}}(\epsilon)+S_{e\hat{\bar{n}} e\hat{n}}(\epsilon)
 +S_{e\hat{n} e\hat{\bar{n}}}(\epsilon)+S_{h \hat{\bar{n}} e \hat{\bar{n}}}(\epsilon)
 ].\label{GS1}
\end{eqnarray}
\end{widetext}
The two equations clearly show the origination of charge and spin conductance.
$ G_c $ is contributed by two kinds of Andreev reflections,\cite{Blonder_Andreev}
while $ G_{s\hat{n}}$ comes from spin-flip reflection \cite{Lv_Spin-flip}
and equal-spin Andreev reflection. \cite{He_AR}

Next, we derive the reflection coefficients $S_{\alpha e\hat{n}}$
($\alpha=e\hat{n}, e\hat{\bar{n}}, h\hat{n}, h\hat{\bar{n}}$)
by using non-equilibrium Green function method.
At the beginning we solve the retarded Green function of
the right-side superconductor and MZM decoupled to the left-side normal lead,
$ {\bf g}_R^r(t) \equiv -i\theta(t) \langle \{  \hat{\psi}_R^T(t), \hat{\psi}_R^{\dagger}(0)  \} \rangle $,
where $ \hat{\psi}_R(t) = (\Sigma_k c_{k\hat{n}}(t), \Sigma_k c_{k\hat{\bar{n}}}(t), \Sigma_k c_{k\hat{n}}^{\dagger}(t), \Sigma_k c_{k\hat{\bar{n}}}^{\dagger}(t), \gamma(t)) $
is in the Nambu representation.\cite{Sun_GS}
By a Fourier transformation the Green function is converted into the energy space
$ {\bf g}_R^r(E) = \int_{-\infty}^{+\infty} dt e^{iEt} {\bf g}_R^r(t) $. Using a wide-band approximation, ${\bf g}_R^r(E) $ is derived to be a $ 5 \times 5 $ matrix\cite{Sun_SC1,Sun_SC2}
\begin{equation*}
	{\bf g}_R^r (E)=-i \pi \rho_S\ \beta (E)
	\begin{pmatrix}
		1 & 0 & 0 & \frac{\Delta}{E}\\
		0 & 1 & -\frac{\Delta}{E} & 0\\
		0 & -\frac{\Delta}{E} & 1 & 0\\
		\frac{\Delta}{E} & 0 & 0 & 1\\
	\end{pmatrix} \bigoplus \frac{1}{E+i0^+},
\end{equation*}\\
where $ \rho_S $ is the density of states of
the s-wave superconductor in the normal states,
$ \beta (E) = E/(i \sqrt{\Delta^2 - E^2}) $ when $ |E|<\Delta $
and $ \beta (E) =  |E|/ \sqrt{E^2 - \Delta^2 } $ when $ |E|>\Delta $.

In the same way, the retarded Green function of the left-side normal lead
decoupled to the right-side s-wave superconductor and MZM
is $ {\bf g}_L^r(E)=-i\pi \rho {\bf I}_{4} $,\cite{Sun_Lead}
where $ \rho $ is density of states of the normal lead
and $ {\bf I}_{4} $ is a $ 4 \times 4 $ unit matrix with the based operators
$\hat{\psi}_L(t) = (\Sigma_k d_{k\hat{n}}(t), \Sigma_k d_{k\hat{\bar{n}}}(t),
\Sigma_k d_{k\hat{n}}^{\dagger}(t), \Sigma_k d_{k\hat{\bar{n}}}^{\dagger}(t)) $.
We set $ \alpha = e\hat{n}, e \hat{\bar{n}}, h\hat{n}, h\hat{\bar{n}} $
indicating the four modes of normal lead,
the spin-$\hat{n}$/$\hat{\bar{n}}$ electron and hole modes.
In order to solve the Green functions for the whole coupling system,
we introduce the retarded self-energy
${\bf\Sigma}_\alpha^r= {\bf H}_{R \alpha} {\bf g}_{L,\alpha\alpha}^r {\bf H}_{\alpha R} $
due to the coupling of the right side to $ \alpha $ mode of the normal lead,
where $ {\bf H}_{\alpha R} $ is the $ 1 \times 5 $
coupling Hamiltonian between the right side and $ \alpha $ mode
and $ {\bf H}_{R\alpha} = {\bf H}_{\alpha R}^{\dagger} $, with
\begin{eqnarray}
H_{e\hat{n} R} &=&(\begin{array}{lllll}
 t_S, & 0,& 0,& 0, & it\cos(\theta/2)
 \end{array}), \nonumber\\
H_{e\hat{\bar{n}} R} &=&(\begin{array}{lllll}
 0, & t_S,& 0,& 0, & it\sin(\theta/2) e^{-i\phi}
 \end{array}), \nonumber\\
H_{h\hat{n} R} &=&(\begin{array}{lllll}
 0, & 0,& -t_S,& 0, & it\cos(\theta/2)
 \end{array}), \nonumber\\
H_{h\hat{\bar{n}} R} &=&(\begin{array}{lllll}
 0, & 0,& 0,& -t_S, & it\sin(\theta/2) e^{i\phi}
 \end{array}). \nonumber
\end{eqnarray}
Then, for the whole coupling system, the retarded Green function of superconductor and MZM
is derived by Dyson equation ${\bf G}_R^r (E)=({{\bf g}_R^r}^{-1}(E)
-\sum_{\alpha}{\bf\Sigma}_{\alpha}^r )^{-1} $,
and the advanced Green function ${\bf G}_R^a (E)=[{\bf G}_R^r (E)]^\dagger $.
At last, the reflection coefficients are\cite{Sun_GS, Cheng_GS}
\begin{equation}
 S_{\alpha e \hat{n}} (E)= Tr({\bf\Gamma}_{e \hat{n}} {\bf G}_R^r  {\bf\Gamma}_\alpha {\bf G}_R^a)
\end{equation}
for $ \alpha \ne e\hat{n} $,
while $ S_{e \hat{n} e \hat{n}} = 1-S_{e \hat{\bar{n}} e \hat{n}}-S_{h \hat{n} e \hat{n}}-S_{h \hat{\bar{n}} e \hat{n}} $,
with the line width function
$ {\bf \Gamma}_\alpha (E)=i[{\bf\Sigma}_\alpha^r
-({\bf\Sigma}_\alpha^r )^\dagger] $.
Similarly, the reflection coefficients $ S_{\alpha e \hat{\bar{n}}} (E)$
can also be obtained.

Usually for a finite superconducting gap $\Delta$,
the analytical expressions of reflection coefficients are complicate.
But by setting $ \Delta \rightarrow \infty $,
the low-energy results keep accurate and are of the much simpler forms below:
\begin{eqnarray}
S_{e \hat{n} e \hat{n}}
 &=& \frac{1}{A} \left[ (E-Ex^4-2 \Gamma x \cos \phi \sin \theta )^2 \right. \nonumber\\
  && \left. +4\Gamma^2 (\sin^2 \frac{\theta}{2}- x^2 \cos^2 \frac{\theta}{2})^2 \right],\label{SS1} \\
S_{e \hat{\bar{n}} e \hat{n}} &= &
 \frac{\Gamma^2}{A}  \left|2x- i (e^{i \phi}-x^2 e^{-i \phi}) \sin \theta\right|^2, \label{SS2} \\
 S_{h \hat{n} e \hat{n}} &=& \frac{4\Gamma^2}{A}  \left|
  \cos^2 \frac{\theta}{2}+x^2 \sin^2 \frac{\theta}{2} e^{2i \phi}\right|^2, \label{SS3} \\
 S_{h \hat{\bar{n}} e \hat{n}} &=&
  \frac{1}{A} \left| 2x[ E(1+x^2) +i \Gamma] \right. \nonumber\\
 && \left. + \Gamma (e^{-i\phi} - x^2 e^{i\phi}) \sin \theta \right|^2, \label{SS4}
\end{eqnarray}
where $ A=(1+x^2 )^2 [4\Gamma^2+E^2 (1+x^2 )^2] $,
$ x=\pi^2 \rho \rho_S t_S^2 $ represents the coupling strength between
the s-wave superconductor and the normal lead,\cite{Hofstetter_x1, Lee_x2}
and $ \Gamma=\pi \rho t^2 $ indicates the coupling strength
between the MZM and the normal lead.
When $\Gamma=0$ (i.e. decoupling to the MZM),
the spin-flip reflection and equal-spin Andreev reflection disappear,
only the normal reflection and normal Andreev reflection occur with
$S_{e \hat{n} e \hat{n}} = \frac{(1-x^2)^2}{(1+x^2)^2}$ and
$S_{h \hat{\bar{n}} e \hat{n}} = \frac{4 x^2}{(1+x^2)^2}$,
because the s-wave superconductor only possesses spin-singlet Cooper pairs.
Here $S_{e \hat{n} e \hat{n}}$ and $S_{h \hat{\bar{n}} e \hat{n}}$
are independent of the energy $E$, due to taking the limit of
$\Delta \rightarrow \infty \gg |E|$.
On the other hand, when $x=0$ (i.e. decoupling to the s-wave superconductor),
four reflection coefficients exist usually with
$S_{h \hat{n} e \hat{n}} = \frac{4\Gamma^2}{A} \cos^4\frac{\theta}{2}$,
$S_{h \hat{\bar{n}} e \hat{n}} =S_{e \hat{\bar{n}} e \hat{n}}= \frac{\Gamma^2}{A} \sin^2\theta$,
and $S_{e \hat{n} e \hat{n}} = \frac{1}{A} (E^2+4\Gamma^2\sin^4\frac{\theta}{2})$. This result is consistent with Ref.[\onlinecite{He_AR}].
In this case, the sum of four Andreev reflection coefficients are
$S_{h \hat{n} e \hat{n}} +S_{h \hat{\bar{n}} e \hat{n}} +
S_{h \hat{n} e \hat{\bar{n}}} +S_{h \hat{\bar{n}} e \hat{\bar{n}}} = \frac{4\Gamma^2}{E^2+4\Gamma^2}$,
i.e. the ZBCCP appears which has extensively been studied in many references.\cite{Mourik_MZME, Das_MZME, Chen_MZME, Nichele_MZME, Zhang_MZME,Lutchyn_MZME, Nadj-Perge_Fe_Chain, Wang_MZM_TS, Xu_MZM_TS1, Xu_MZM_TS2, Sun_MZM_TS3, Wang_iron-base2, Law_ZBP, Zhu_iron-base3, Vaitiekenas_MZME}

In the following, we will focus on two cases,
the finite superconducting gap $\Delta$ and $\Delta \rightarrow \infty$.
For the finite $\Delta$, we take $\Delta=1$ as the unit of energy.
Then the unit of $\Gamma$ the coupling strength between the normal lead and MZM,
$E$ the energy of the incident electron and $k_B T$ the temperature is $\Delta$.
On the other hand, when $\Delta$ is much larger than the energy $E$
(e.g. $|E|<0.3|\Delta|$), we set $\Delta \rightarrow \infty$.
In this limit $\Delta \rightarrow \infty$,
the expressions of the reflection coefficients and conductances
are in very simple forms.

\begin{figure}
	\includegraphics[width=\columnwidth]{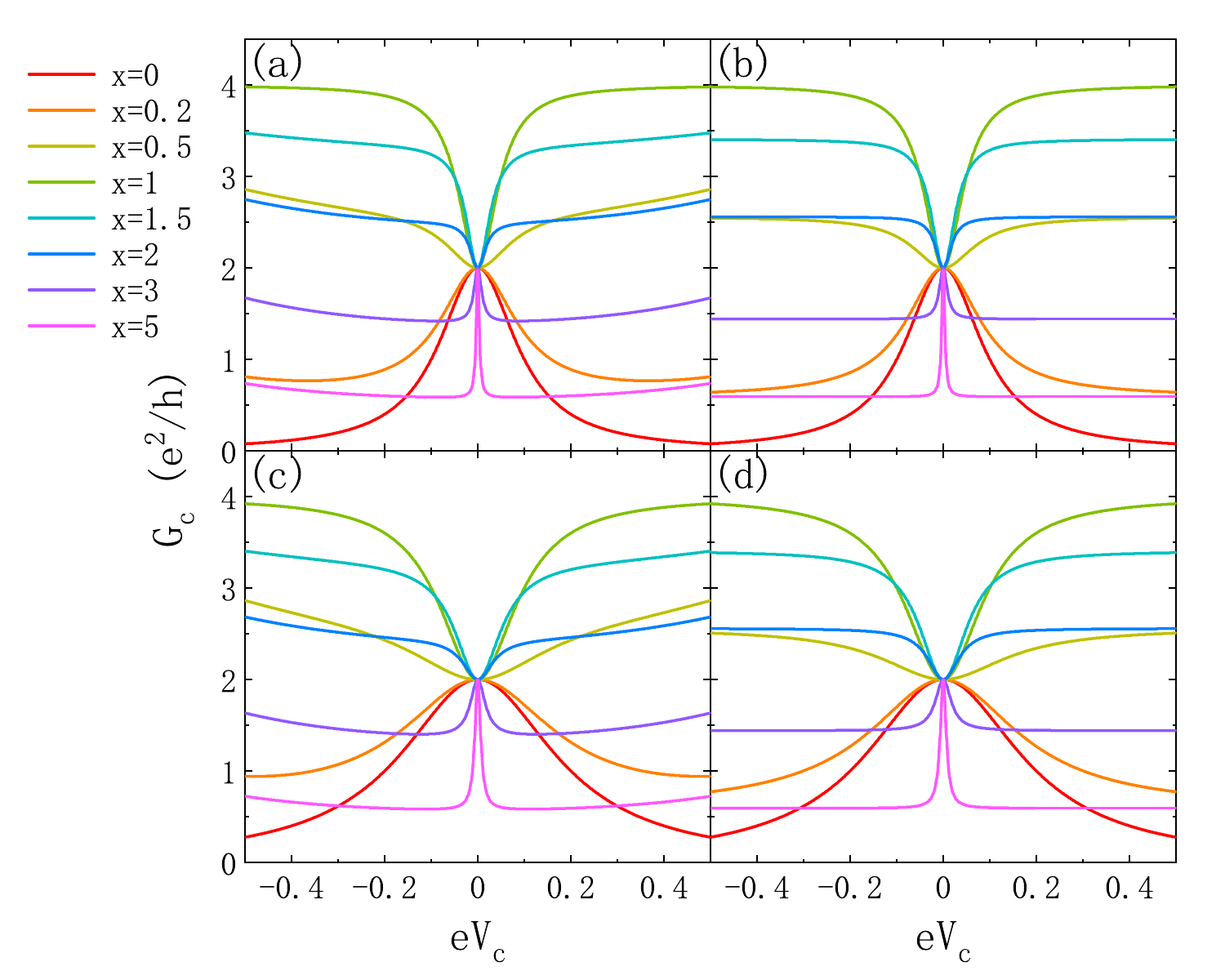}%
	\caption{\label{FIG2} The charge conductance
$ G_c $ as a function of $ V_c $ with (a)(c) $ \Delta=1 $ and (b)(d) $ \Delta \rightarrow \infty $. Parameter $\Gamma$=0.05 for (a)(b), and $\Gamma$=0.1 for (c)(d).
The temperature $k_B T=0$.}
\end{figure}

\section{\label{SEC3} Results of charge transport}
We first study how the charge conductance $G_c$ and ZBCCP are affected by the coupling
of the s-wave superconductor.
Fig.~\ref{FIG2} shows $G_c$ versus the charge bias $V_c$
for both $ \Delta=1 $ and $ \Delta \rightarrow \infty $.
Here the parameters in Fig.~\ref{FIG2}(a,c) and
Fig.~\ref{FIG2}(b,d) are the same, except $\Delta=1$ in (a,c)
and $ \Delta \rightarrow \infty $ in (b,d),
and one can regard $ \Gamma $ as the characteristic energy when comparing
results in two regimes of $ \Delta $.
Here $G_c$ is independent of how the spin direction $\hat{n}$ of the incident electron
is taken in the calculation, because charge bias is a scalar.
When $|eV_c|<0.3$, the conductances of the finite gap $\Delta=1$ are quite consistent
with those of $ \Delta \rightarrow \infty $,
so the $ \Delta \rightarrow \infty $ approximation
used to get Eqs.(\ref{SS1}-\ref{SS4}) is appropriate,
and we can regard the results of
$ \Delta \rightarrow \infty $ just as $ \Delta = 1 $ regime.
By substituting Andreev reflection coefficients in Eqs.(\ref{SS3}-\ref{SS4}) into
Eq.(\ref{GC1}), the conductance $G_c$ at $ \Delta \rightarrow \infty $ and $k_B T=0$
can be analytically obtained:
\begin{equation}
	G_c=\frac{2e^2}{h} \frac{4\Gamma^2+8(e V_c)^2 x^2}{ 4\Gamma^2+(e V_c)^2 (1+x^2 )^2} \label{GC2},
\end{equation}
If the normal lead is decoupled to the MZM ($\Gamma=0$),
$G_c = \frac{2e^2}{h}\frac{8x^2}{(1+x^2)^2}$,
which reaches the maximum $\frac{4e^2}{h}$ at $x=1$
(the perfect coupling between normal lead and s-wave superconductor).
Notice that the coupling strength $ x \equiv \pi^2 \rho \rho_S t_S^2 $ comes from
the hopping strength $t_S$ and the densities of states $\rho$ and $\rho_S$
in normal lead and s-wave superconductor.
Only when $\rho$, $\rho_S$ and $t_S$ match very well (i.e. at $x=1$),
the perfect resonant Andreev reflection happens
and then the charge conductance $ G_c $ reaches the maximum ${4e^2}/{h}$.
If $x<1$ or $x>1$, the resonance is broken and $G_c$ will be less than ${4e^2}/{h}$,
which is also shown in Refs.[\onlinecite{Hofstetter_x1, Beenakker_NS_junction, Claughton_NS}].
On the other hand, if decoupled to s-wave superconductor ($x=0$),
$ G_c= \frac{2e^2}{h}\frac{4\Gamma^2}{4\Gamma^2+(eV_c)^2} $,
which is a Lorenz ZBCCP with the peak height ${2e^2}/{h}$
consistent with literature.\cite{Law_ZBP}
When both $\Gamma$ and $x$ are non-zero, the incident electron can be Andreev reflected
as a hole by both the MZM and s-wave superconductor,
and we can see an essential role played by the competition between them.
For a small coupling strength $x$ to s-wave superconductor,
the conductance is still dominated by the MZM and
the shape of $ G_c-V_c $ curve retains a ZBCCP
(see the curves with $ x=0, 0.2 $ in Fig.~\ref{FIG2}).
When $x $ is enlarged, the $ G_c $ contribution from
s-wave superconductor is not negligible.
Then the non-zero-bias $ G_c $ could be enhanced by the superconductor
and change the shape of curve from a ZBCCP to a zero-bias valley
(see the curves with $ x=0.5, 1, 1.5, 2 $ in Fig.~\ref{FIG2}).
This means that a MZM doesn't certainly induce a ZBCCP,
a zero-bias valley is also a possible result.
Because the normal lead is inevitably coupled to the s-wave superconductor,
even if the ZBCCP is not observed in the experiment,
it does not indicate that the MZM does not exist.
What's more, the charge conductance $G_c$ at zero bias is always
the quantized value $ 2e^2/h$ in both ZBCCP and zero-bias valley regimes,
because the MZM induces a resonant Andreev tunneling at zero energy
and the contribution of the s-wave superconductor is completely suppressed.
This is similar to a circuit where two resistors $R_1$ and $R_2$
are connected in parallel, with $R_1$ being nonzero and $R_2=0$,
and the current completely passes through $R_2$ and the
contribution of $R_1$ is completely suppressed.
With the further increase of $x$, the non-zero-bias charge conductance
reduces because of the mismatching of density of states
between the normal lead and superconductor,\cite{Beenakker_NS_junction}
so that the $ G_c-V_c $ curve returns to a ZBCCP again
(see the curves with $ x=3, 5 $ in Fig.~\ref{FIG2}).
The influence by parameter $ \Gamma $ is limited.
It only impacts the width of the peaks and valleys.

\begin{figure*}
	\includegraphics[width=2\columnwidth]{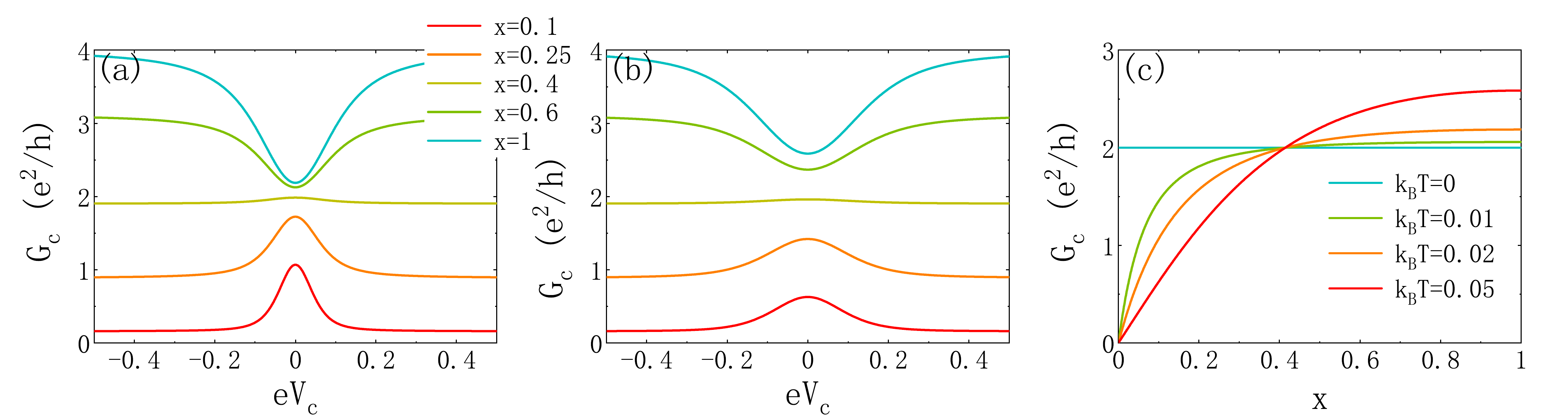}%
	\caption{\label{FIG3}
(a) and (b) The charge conductance $ G_c $ as a function of $ V_c $
for the different coupling strengths with the finite temperatures
(a) $ k_B T=0.02 $ and (b) $ k_B T=0.05 $.
(c) The zero-bias charge conductance $ G_c $ as a function of the coupling strength $ x $.
Parameters $ \Delta \rightarrow \infty $ and $\Gamma= 0.1x $.}
\end{figure*}

In order to better compare with the experiment,
we also study the non-zero temperature case.
In addition, we set that the coupling strength $x$ is proportional to
the coupling strength $\Gamma$,
because that both of them are simultaneously regulated by the gate voltage
or the distance between sample and probe
in the experiments,\cite{Mourik_MZME,Zhu_iron-base3} or theoretically adjusted
by a tunnel barrier between lead and nanowire.\cite{Vuik_ZBV}
The charge conductance $ G_c = d (eI_{\hat{n}}+eI_{\hat{\bar{n}}}) / d V_c $
where the particle currents $I_{\hat{n}}$ and $I_{\hat{\bar{n}}}$
can be obtained from Eq.(\ref{In}).
Figs.~\ref{FIG3} (a) and (b) show $G_c$ versus
the charge bias $V_c$ for the different coupling strength with $\Gamma=0.1x$.
For small coupling strengths, the curve $G_c$-$V_c$ exhibits
a zero-bias peak (see the curves with $\Gamma=0.1x=0.01$ and $0.025$).
With the increase of the coupling strengths, the zero-bias peak gradually evolves
to the zero-bias valley, which is similar to that in Fig.~\ref{FIG2}.
However, due to the effect of the finite temperature,
the height of the zero-bias peak at the small coupling strengths is slightly
less than $2e^2/h$ and the bottom of the zero-bias
valley at the large coupling strengths is higher than $2e^2/h$.
These results, the evolvement from zero-bias peak to zero-bias valley,
are well consistent with the recent experiment.\cite{Vaitiekenas_MZME}
A recent theoretical work by Vuik {\sl et al.}\cite{Vuik_ZBV}
studied the normal lead-superconducting nanowire junction
with the existence of the spin-orbit interaction and the Zeeman energy,
and they have shown a similar peak-to-valley evolvement with the decrease of the tunnel barrier.
Here we exhibit that this peak-to-valley evolvement
derives from the coupling of the s-wave component of the superconductor.

Fig.~\ref{FIG3}(c) shows the zero-bias charge conductance $G_c$
versus coupling strength $x$ with $\Gamma=0.1x$ for the different temperature.
At the zero temperature, $G_c$ is determined by zero-energy reflection coefficients
and always keeps the value $ 2e^2/h $ for the dominance of MZM.
With the increase of the temperature,
$ G_c $ decreases at the small coupling strength and it increases
at the large coupling strength.
For a fixed temperature, the charge conductance
$G_c$ monotonously increases from $0$ through $2e^2/h$ to more than $2e^2/h$
with the increase of the coupling strengths,
and it shows a plateau-like shape with the plateau value being about $2e^2/h$,
which is in good agreement with the recent experiment.\cite{Zhu_iron-base3}

\section{\label{SEC4} Results of spin transport}

\begin{figure}
	\includegraphics[width=\columnwidth]{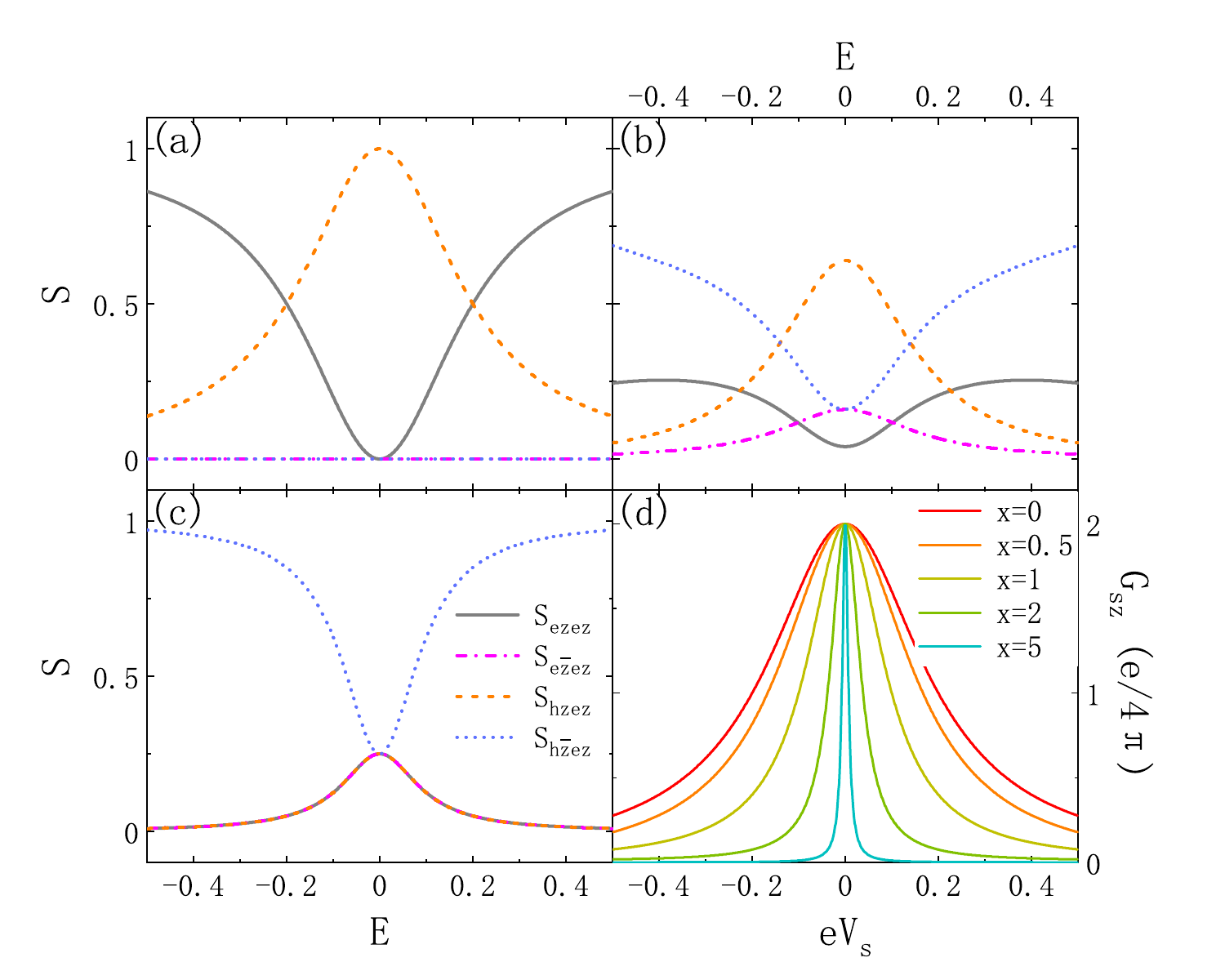}%
	\caption{\label{FIG4} (a-c) Reflection coefficients as a function of
energy $E$ for the incident $+z$-spin electron with
the coupling strength $ x $=0 (a), 0.5 (b), and 1 (c).
(d) Spin conductance as a function of spin bias at $k_B T=0$.
The parameters $ \Delta=1 $ and $ \Gamma=0.1 $.
}
\end{figure}

Next, we focus on the spin transport.
In contrast to charge transport, we expect that the spin transport is not affected
by the coupling of s-wave superconductor.
This is because that s-wave superconductor with the spin-singlet Cooper pairs
is insulator for spin current.
Fig.~\ref{FIG4}(a-c) shows the reflection coefficients $ S_{\alpha e z}$-$E $ curves
with different values of $ x $
for the incident electron's spin at the $+z$ direction
($\theta=\phi=0$) and $\Delta=1$.
These results are almost the same as those in the limit $ \Delta \rightarrow \infty $.
When normal lead is decoupled to the s-wave superconductor ($x=0$),
there are only two processes: normal reflection and equal-spin Andreev reflection.
The spin-flip reflection and normal Andreev reflection disappear because
the MZM only couples to the $+z$-spin electron.\cite{He_AR}
If the incident energy $ E=0 $, $ S_{h z e z} =1 $,
indicating that there is a complete equal-spin Andreev reflection,
which is consistent with Ref.[\onlinecite{He_AR}].
As $ x=0 $, the electron with $ -z $ spin is completely normal reflected,
so charge and spin transport are determined by $ S_{h z e z} $ process,
and this curve has the same shape as $ x=0 $ curves
in Fig.~\ref{FIG2}(c) and Fig.~\ref{FIG4}(d).
With the increase of coupling strength $x$ from 0,
the normal Andreev reflection $ S_{h \bar{z} e z} $ appears and increases
due to the reflection by s-wave superconductor,
meanwhile $ S_{e z e z} $ and $ S_{h z e z} $ are suppressed
[see Fig.~\ref{FIG4}(b)(c)].
Here the spin-flip reflection $ S_{e \bar{z} e z} $ also appears
due to the combination of both s-wave superconductor and MZM.
When $ x=1 $, the coupling between the normal lead and s-wave superconductor is complete,
then the normal Andreev reflection $ S_{h \bar{z} e z} $ is major,
while the other three coefficients are small
with curves coinciding together [see Fig.~\ref{FIG4}(c)].
As shown in Eqs.[\ref{GC1},\ref{GS1}], normal Andreev reflection
contributes to charge transport, but doesn't lead to spin transport.
So the enhancement of $ S_{h \bar{z} e z} $ increases $ G_c $
but decreases $ G_s $ at non-zero-bias.
In addition, all reflection coefficients are symmetric about incident energy $E$,
i.e. $S_{\alpha'\alpha}(-E)=S_{\alpha'\alpha}(E)$.

Fig.~\ref{FIG4}(d) shows the spin conductance $G_{sz}$ versus the
$ +z $-direction spin bias $V_s$ with $ \Delta=1 $ and $ \Gamma=0.1$.
When decoupled to the s-wave superconductor ($x=0$),
the $G_{sz}$-$V_s$ curve shows a zero bias peak
with a half-height width $2\Gamma$.
The peak height is quantized at the value $ e/2\pi $ at $V_s=0$.
In particular, no matter whether normal lead is coupled to superconductor
and how strong the coupling is,
this zero-bias peak of $G_{sz}$ can well survive and the peak height remains unchanged.
The participation of s-wave superconductor cuts down the width of the peak.
This is essentially different from the charge conductance,
in which the ZBCCP varies into the zero-bias valley with the increase of $x$ from $0$ to $1$.
The results of $G_{sz}$ at $|eV_s| < \Delta$ for finite $\Delta$ are
nearly the same as those of $ \Delta \rightarrow \infty $.
The spin conductance $G_{sz}$ at $ \Delta \rightarrow \infty $ can be analytically obtained
by combining the Eqs.(\ref{GS1}), (\ref{SS2}), and (\ref{SS3}),
\begin{equation}
G_{sz}=\frac{e}{2\pi} \frac{4\Gamma^2}{ 4\Gamma^2+(e V_s)^2 (1+x^2 )^2} \label{GS2}.
\end{equation}
Eq.(\ref{GS2}) shows that $ G_{sz}$-$V_s $ curve is a Lorentz-shape peak.
In particular, the height of the zero-bias peak always holds at $ e/2\pi $
regardless of the systemic parameters ($x$ and $\Gamma$),
and the half width at half maximum is $ 2\Gamma/(1+x^2) $.

Let us explain why zero-bias spin conductance peaks always exist.
On one hand, the MZM serves as a zero-energy resonance state,
it dominates the transport properties at zero bias.
On the other hand, the MZM has functions similar to p-wave superconductor,
and it gives rise to the equal-spin Andreev reflection,
which can contribute to both spin transport and charge transport.
By combining with the above two reasons, it results in
the zero-bias spin conductance and charge conductance
having the robust values $ e/2\pi $ and $2e^2/h$, respectively.
For the s-wave superconductor,
its Cooper pairs are of charge $ 2e $ and spin singlet.
They transit charge but no spin.
So the coupling of the s-wave superconductor
may enhance the charge conductance and make the ZBCCP turn into a zero-bias valley,
but it always suppresses the spin conductance at non zero bias
and reduces the width of zero-bias spin conductance peak,
leading to the survival of this peak.
Because the spin conductance curve keeps a peak,
experimentally we could detect MZMs by measuring spin conductance
and rule out the misleading zero-bias valley in charge transport.

\begin{figure}
	\includegraphics[width=\columnwidth]{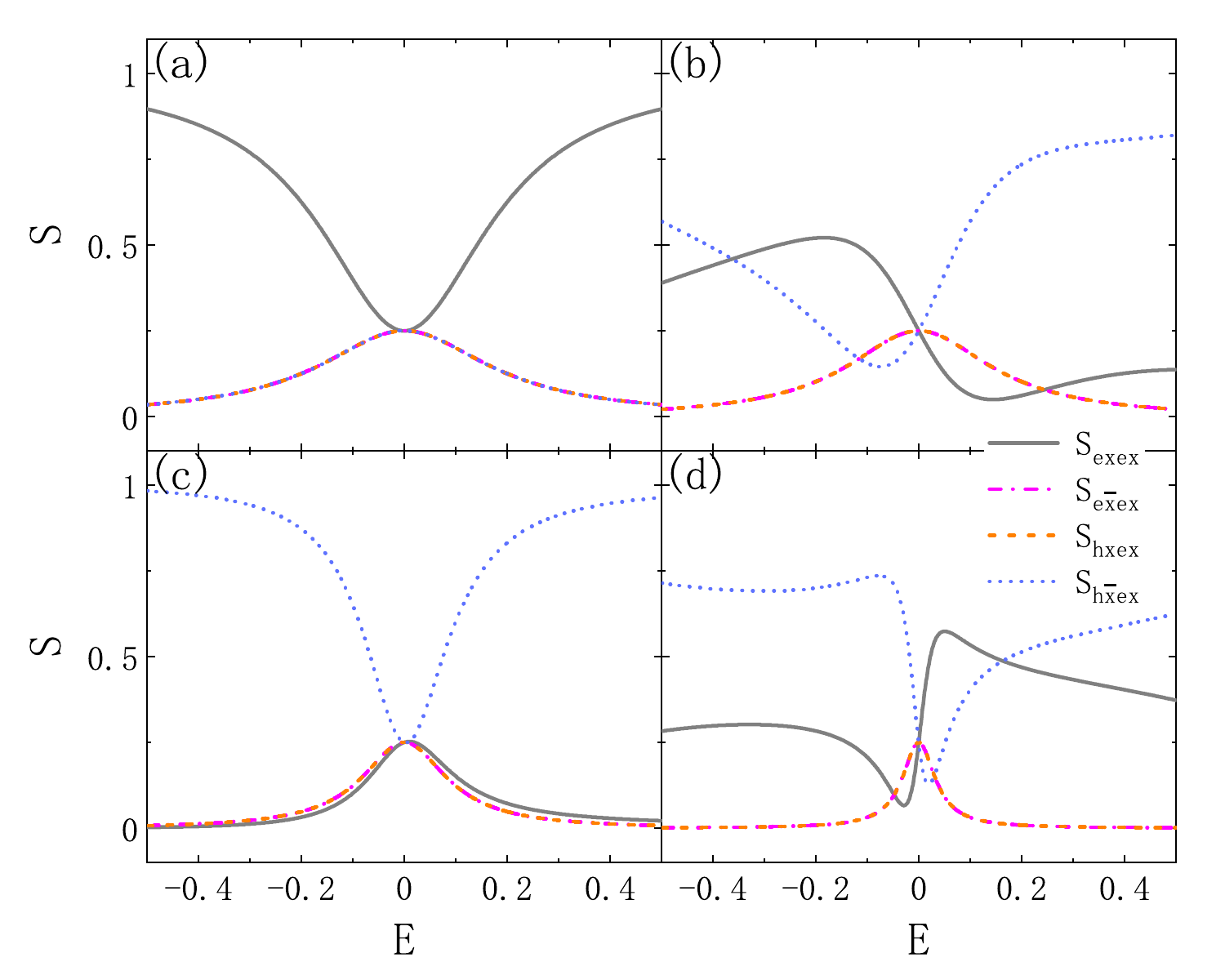}%
	\caption{\label{FIG5}
Reflection coefficients as a function of energy $E$ for
the incident $+x$-spin electron with $ \Delta=1 $ and $ \Gamma=0.1 $.
The coupling strength $ x $=0 (a), 0.5 (b), 1 (c), and 2 (d).}
\end{figure}

Different from charge transport, spin is a vector, so spin bias is also a vector.\cite{Shi_Spin_current, Sun_Spin_current}
Below, we study the spin transport under $+x$-direction spin bias.
Fig.~\ref{FIG5} shows the four reflection coefficients for
an incident $+x$-spin electron with different coupling strength $x$.
The results are distinctly different from those in $+z$ direction
(see Figs.~\ref{FIG4} and \ref{FIG5}).
When $x=0$, four reflection coefficients $S_{\alpha ex}$ for the incident $+x$-spin electron
are totally non-zero [Fig.~\ref{FIG5}(a)],
but for the incident $+z$-spin electron $S_{e\bar{z} ez}$ and $S_{h\bar{z} ez}$
are exactly zero [Fig.~\ref{FIG4}(a)].
For $ x\ne 0 $, normal reflection and normal Andreev reflection coefficients
are asymmetrical with $ E=0$,
while spin-flip reflection and equal-spin Andreev reflection coefficients keep symmetrical.
This phenomenon is evident especially for $ x=0.5$ and $2$ (see Fig.~\ref{FIG5}(b) and (d)).
Both $ S_{e x e x} (E) $ and $ S_{h \bar{x} e x} (E) $
have a peak and a valley in the opposite sides of $ E=0 $,
which is a typical Fano resonance.\cite{Ricco_Fano1,Schuray_Fano2}
However, for an incident $+z$-spin electron, four reflection coefficients
are symmetrical with $E=0$
and without the Fano resonance shape at all.

Let us discuss the origin of the Fano resonance
and the difference between the incident $+z$-spin and $+x$-spin electrons. We first consider Andreev reflections.
For the incident $+z$-spin electron,
because MZM is only coupled to this spin,\cite{He_AR}
it only leads to equal-spin Andreev reflection $ S_{h z e z}$.
In addition, s-wave superconductor just induces a normal Andreev reflection.
So there is no coherent and Fano resonance does not happen.
Unlike $+z$-spin electron, the incident $+x$-spin electron can be decomposed
to $ |ex \rangle=\frac{1}{\sqrt{2}}(|ez\rangle+|e\bar{z}\rangle) $.
In the presence of only MZM, the $ |e\bar{z}\rangle $ component
has no contribution to Andreev reflections,
while the $ |ez\rangle $ component is equal-spin Andreev reflected into $ |hz\rangle $,
which is recomposed to $ |hz\rangle=\frac{1}{\sqrt{2}}(|hx\rangle+|h\bar{x}\rangle) $.
It explains that both normal and equal-spin Andreev reflections could happen
to an incident $+x$-spin electron at $x=0$ (see Fig.~\ref{FIG5}(a)).
Meanwhile, the $+x$-spin electron can be still reflected into
an opposite-spin hole by the s-wave superconductor.
As a result, the reflected $-x$-spin hole has both a resonant spectroscopy from MZM
and a continuous spectroscopy from s-wave superconductor.
The coherent addition induces the Fano resonance of $ S_{h \bar{x} e x} (E) $.
On the contrary, the $+x$-spin hole comes from only the MZM,
and so the $S_{h x e x}$-$E$ curve keeps symmetrical with $ E=0 $.
Similarly, we can explain why normal reflection appears Fano resonance
but spin-flip reflection does not.

Next, we study the spin conductance under the spin bias of any direction.
For the gap $ \Delta=1$, the spin conductance curve under
$+x$-direction spin bias ($ \theta= \pi/2$ and $\phi = 0 $)
is almost the same as that of $+z$-direction,
although the reflection coefficients are very different.
It also robustly holds a zero-bias spin conductance peak with the peak height $ e/2\pi $
and keeps the peak which can only be narrowed by the coupling strength $x$.
Actually, for the spin bias of any direction, the spin conductance $G_{s\hat{n}}$ is also
almost the same.
Furthermore, in the $ \Delta \rightarrow \infty $ limit,
$ G_{s\hat{n}}$-$V_s$ curve is exactly the same as that shown in Eq.(\ref{GS2})
regardless of the angles $\theta$ and $\phi$ of the spin bias,
appearing the shape of Lorentz peak with a half-height width $ 2\Gamma/(1+x^2) $.
It is worth mentioning that in ordinary spin-triplet superconductors,
spin conductance usually depends on the direction of spin bias.
The isotropy of spin bias provides convenience for experimental measurements.
At last, let us discuss how to detect the spin conductance.
Up to now, the spin current has been generated
by using various methods in the experiments,
e.g. by using the ferromagnetic electrodes,\cite{Valenzuela_Spin_detect}
spin Hall effect,\cite{Cornelissen1022} and spin Seebeck effect.\cite{Hirobe30}
The spin current can be detected by the ferromagnetic electrodes,\cite{Han_Spin_current}
the inverse spin Hall effect,\cite{Valenzuela_Spin_detect,Li70}
the induced electric field,\cite{Sun_Spin_current} etc.
The spin bias could be measured by a large open quantum dot
or double quantum dot.\cite{Koop_Spin_detect,Sun_Spin_bias}
So the spin conductance should be measurable in the present technology.

\section{\label{SEC5} Conclusion}

In summary, the transport through a normal lead coupled to a MZM and
an s-wave superconductor is studied.
We show that the charge (spin) conductance is the quantized
value $ 2e^2/h $ ($ e/2\pi $) at zero charge (spin) bias.
In the presence of s-wave superconductor, the ZBCCP could be changed to a zero-bias valley.
Since the coupling to s-wave superconductor is experimentally inevitable,
no ZBCCP does not mean no MZM.
However, spin conductance always keeps a zero-bias peak.
So measuring spin transport properties could be a more reliable method
to judge the existence of MZMs.
Furthermore, the reflection coefficients are dependent of
spin direction of the incident electron.
For certain spin, the coefficients may appear a Fano resonance shape.

\section*{Acknowledgement}
This work was financially supported by National Key R and D Program of China (Grant No. 2017YFA0303301), NSF-China (Grant No. 11921005), the Strategic Priority Research Program of Chinese Academy of Sciences (Grant No. DB28000000), and Beijing Municipal Science \& Technology Commission (Grant No. Z191100007219013).


\end{document}